\documentclass[preprint]{aastex}
\begin{document}

\title{The Great Observatories All-Sky LIRG Survey: Comparison of Ultraviolet and 
Far-Infrared Properties}
\author{
Justin H. Howell,\altaffilmark{1}
Lee Armus,\altaffilmark{1}
Joseph M. Mazzarella,\altaffilmark{2}
Aaron S. Evans,\altaffilmark{3,4}
Jason A. Surace,\altaffilmark{1}
David B. Sanders,\altaffilmark{5}
Andreea Petric,\altaffilmark{1}
Phil Appleton,\altaffilmark{6}
Greg Bothun,\altaffilmark{7}
Carrie Bridge,\altaffilmark{1,11}
Ben H.P. Chan,\altaffilmark{2}
Vassilis Charmandaris,\altaffilmark{8,9}
David T. Frayer,\altaffilmark{6}
Sebastian Haan,\altaffilmark{1}
Hanae Inami,\altaffilmark{1}
Dong-Chan Kim,\altaffilmark{3}
Steven Lord,\altaffilmark{6}
Barry F. Madore,\altaffilmark{2,10}
Jason Melbourne,\altaffilmark{11}
Bernhard Schulz,\altaffilmark{6}
Vivian U,\altaffilmark{5}
Tatjana Vavilkin,\altaffilmark{12}
Sylvain Veilleux,\altaffilmark{13}
Kevin Xu\altaffilmark{6}
}
\altaffiltext{1}{Spitzer Science Center, MS 220-6, California Institute 
of Technology, Pasadena,
CA 91125; jhhowell@ipac.caltech.edu}
\altaffiltext{2}{Infrared Processing \& Analysis Center, MS 100-22, 
California Institute of Technology,
Pasadena, CA 91125}
\altaffiltext{3}{Department of Astronomy, University of Virginia, P.O. Box 400325, Charlottesville, VA 22904}
\altaffiltext{4}{National Radio Astronomy Observatory, 520 Edgemont Road, Charlottesville, VA 22903}
\altaffiltext{5}{Institute for Astronomy, University of Hawaii, 2680
Woodlawn Drive, Honolulu, HI 96822}
\altaffiltext{6}{NASA Herschel Science Center, IPAC, MS 100-22, 
California Institute of Technology,
Pasadena, CA 91125}
\altaffiltext{7}{Department of Physics, University of Oregon, Eugene, OR 97403}
\altaffiltext{8}{University of Crete, Department of Physics, Heraklion 71003, 
Greece}
\altaffiltext{9}{IESL/Foundation for Research and Technology - Hellas, 
GR-71110, Heraklion, Greece and Chercheur Associ\'e, Observatoire de 
Paris, F-75014, Paris, France}
\altaffiltext{10}{The Observatories, Carnegie Institution of Washington,
813 Santa Barbara Street, Pasadena, CA 91101}
\altaffiltext{11}{Caltech Optical Observatories, Division of Physics, Mathematics and 
Astronomy, Mail Stop 320-47, California Institute of Technology, Pasadena, CA 
91125}
\altaffiltext{12}{Department of Physics and Astronomy,
State University of New York at Stony Brook, Stony Brook, NY 11794-3800}
\altaffiltext{13}{Department of Astronomy, University of Maryland, College Park, MD 20742}

\begin{abstract}

The Great Observatories All-sky LIRG Survey (GOALS) consists of a 
complete sample of 202 Luminous Infrared Galaxies (LIRGs)
selected from the IRAS Revised Bright Galaxy Sample (RBGS).  The galaxies
span the full range of interaction stages, from isolated galaxies to interacting pairs 
to late stage mergers.  We present a comparison of the UV and infrared properties of 
135 galaxies in GOALS observed by GALEX and Spitzer.  For interacting 
galaxies with separations greater than the resolution of GALEX and Spitzer
($\sim2-6^{\prime\prime}$), we assess the UV and IR properties of each galaxy individually.
The contribution of the FUV to the measured SFR ranges from 0.2\% to 17.9\%, with a 
median of 2.8\% and a mean of $4.0\pm0.4$\%. 
The specific star formation rate of the GOALS sample is extremely high, 
with a median value ($3.9\times10^{-10}~{\rm yr}^{-1}$) that is comparable to the highest 
specific star formation rates 
seen in the Spitzer Infrared Nearby Galaxies Survey sample.  We examine the position 
of each galaxy on the IR excess--UV slope (IRX-$\beta$) diagram as a function of galaxy 
properties, including IR luminosity and interaction stage.  The LIRGs on average have greater IR 
excesses than would be expected based on their UV colors if they obeyed the same 
relations as starbursts with $L_{IR}<10^{11}L_{\odot}$ or normal late-type galaxies.  
The ratio of $L_{IR}$ to the value one would estimate from the 
IRX-$\beta$ relation published for lower luminosity starburst galaxies ranges from 
0.2 to 68, with a median value of 2.7.  A minimum of $19\%$ of the total IR luminosity 
in the RBGS is produced in LIRGs and ULIRGs with red UV colors ($\beta>0$).  
Among resolved interacting systems, $32\%$ contain one galaxy which 
dominates the IR emission while the companion dominates the UV emission.
Only $21\%$ of the resolved systems contain a single galaxy which dominates 
both wavelengths.  

\end{abstract}

\keywords{ultraviolet: galaxies, infrared: galaxies}

\section{Introduction}

The Infrared Astronomical Satellite (IRAS) provided the first unbiased 
survey of the sky at mid and far-infrared wavelengths, giving us a 
comprehensive census of the infrared emission properties of galaxies in the 
local Universe.  A major result of this survey was the discovery of a 
large population of luminous infrared galaxies (LIRGs) which emit a 
large majority of their bolometric luminosity in the far-infrared, 
and have $10^{11}\leq{\rm L_{IR}} [8-1000\mu{\rm m}] <10^{12} L_{\odot}$.  LIRGs are a 
mixture of single galaxies, disk galaxy pairs, interacting systems, and 
advanced mergers.  They exhibit enhanced star-formation rates and a higher 
fraction of Active Galactic Nuclei (AGN) compared to less luminous and 
non-interacting galaxies (Sanders \& Mirabel 1996 and references therein).  
At the highest luminosities, ultraluminous infrared galaxies (ULIRGs: 
${\rm L_{IR}} \geq 10^{12} L_{\odot}$) may represent an important 
evolutionary stage in the formation of QSOs \citep{sanders88a,sanders88b}
and massive ellipticals \citep[e.g.,][]{genzel01,tacconi02}.
Since LIRGs comprise the bulk of the cosmic infrared background and dominate 
the star-formation activity between $0.5 < z < 1$ \citep{lefloch05,caputi06},
they may also play a key role in our understanding of 
the general evolution of galaxies and black holes \citep[e.g.,][]{magorrian98}.

The Great Observatories All-sky LIRG Survey \citep[GOALS;][]{armus09} contains a
complete sample of low-redshift LIRGs and ULIRGs with observations across the 
electromagnetic spectrum.  The GOALS targets are drawn from the IRAS 
Revised Bright Galaxy Sample (RBGS; Sanders et al. 2003), a complete sample 
of 629 galaxies with IRAS $60~\mu$m flux densities ${\rm S}_{60} > 5.24$~Jy, covering the full 
sky above Galactic latitudes $|b| > 5$ degrees.  The 629 galaxies have a 
median redshift of $z = 0.008$ and a maximum redshift of $z=0.088$.  
There are 181 LIRGs and 21 ULIRGs in the RBGS, and these galaxies define 
the GOALS sample.

In LIRGs and ULIRGs, UV radiation is produced by young stars and AGN.   A fraction of the  
UV radiation is absorbed by dust and re-radiated in the far-infrared.  To 
understand the power sources in these galaxies, it is essential to fully 
characterize the energy budget by measuring both the emerging UV and the 
infrared flux.  The relationship between the UV continuum slope and the 
infrared excess (the IRX-$\beta$ correlation) provides a useful parameterization 
of this energy budget.  \citet{charlot00} showed that the IRX-$\beta$ 
relation is a sequence in effective optical depth for star forming systems.  
However this relation does not hold in all systems.  While lower luminosity 
starbursts follow the correlation, ULIRGs do not \citep*{meurer,goldader}.
The GOALS sample allows us to explore the IRX-$\beta$ correlation precisely 
over the luminosity range where it breaks down.  
A detailed study of LIRGs may indicate the luminosity threshold 
or the time during the merger when the UV slope becomes decoupled from 
the IR emission.  Being a flux limited sample of the nearest and most well studied 
LIRGs and ULIRGs, GOALS provides an important local benchmark against which to compare 
the observed visual properties of high redshift galaxies.  This paper looks at global UV 
and IR properties.  Future work will address nearby spatially resolved LIRGs.

This paper is divided into five sections.  The data are discussed in 
\S~2.  Analysis of the sample is presented in \S~3, results are discussed
in \S~4, and conclusions are given in \S~5.  A cosmology of 
$\Omega_{\Lambda}=0.72$, $\Omega_m = 0.28$, with 
$H_0 = 70~{\rm km~s^{-1}~Mpc^{-1}}$ is 
adopted throughout.

\section{Observations and Data Reduction}

The GOALS GALEX sample consists of 135 systems observed as part of 
GALEX Cycle~1 program \#13 (PI Mazzarella), GALEX Cycle~5 program \#38 (PI Howell), 
the Nearby Galaxy Survey (NGS), and the All Sky Survey (AIS).  All systems have been 
observed in both the FUV ($\lambda_{\rm eff} = 1528$\AA) and NUV 
($\lambda_{\rm eff} = 2271$\AA).  Integration times range from $\sim100$s for the 
AIS data to $>1500$s for the Cycle~1, Cycle~5, and NGS data.  Aside from a handful 
of galaxies not yet observed from the Cycle~5 program, the 135 systems described here 
represent all GOALS targets accessible to GALEX.

Photometry was performed on the standard GALEX pipeline data products. 
Since GALEX backgrounds are very low, especially for FUV images, 
standard photometry codes often return a background value of zero.
To accurately measure the background in these images, we followed the 
prescription of \citet[][hereafter GDP]{gildepaz07} using software written by 
those authors and made available to us.  Standard IDL aperture 
photometry codes were used to measure the total UV fluxes inside large 
apertures (typically $1^{\prime}$ radius) matched to Spitzer $24~\mu$m 
MIPS photometry \citep{mazz}.  Aperture centers were taken from \citet{armus09}.  
The resultant UV GALEX photometry of the sample is presented in Table~1.  
In the case of widely separated pairs, only the more IR-luminous component is listed.
Close pairs are treated as a single system, with the combined flux density listed 
in the table.

To test the accuracy of our measurements 
and to ensure meaningful comparisons with published data sets such as GDP, 
galaxies with D25 ellipses \citep{rc3} were measured in D25 elliptical 
apertures.  Little difference was found between the fluxes measured in 
the D25 aperture as compared to the circular aperture.  The seven targets 
in common with the sample of GDP revealed a systematic shift in the 
photometric calibration between different 
versions of the GALEX data reduction pipeline.  To account for this, 
the GDP fluxes have been adjusted for purposes of comparison with the 
GOALS sample.  The raw count rates (before background subtraction) have been 
multiplied by factors of 0.89 (FUV) and 1.05 (NUV).  Fluxes and magnitudes 
were then recalculated after background subtraction.

The resolution of Spitzer allows many interacting pairs or groups to 
be resolved into their component galaxies in the IR.  For systems with 
separations greater than $0.5^{\prime}$, the $70~\mu$m flux ratio was 
used to estimate the fraction of the IRAS $L_{IR}$ coming from each 
galaxy.  Similarly, the $24~\mu$m flux ratio was used for systems 
separated by $0.12^{\prime}<d<0.5^{\prime}$, and systems 
which saturated at $70~\mu$m.  The latter method is inaccurate for 
systems in which the two galaxies have different far-IR colors, such as 
the Arp~299 (NGC3690/IC0694) system \citep*[see][]{char02}.
A total of 93 galaxies in 44 GOALS systems have been resolved in one or 
both GALEX FUV and NUV images.  Photometry of the resolved sources is 
presented in Table~2.

\section{Results}
\subsection{UV Luminosities and Spectral Slopes}

Although selected to be IR luminous, the GOALS sample spans a wide range 
of UV luminosities.  The FUV flux densities range from $2.4\times10^{-16}$ to 
$2.9\times10^{-13}~{\rm erg~s^{-1} cm^{-2} \AA^{-1}}$, with a 
median of $7.3\times10^{-15}~{\rm erg~s^{-1} cm^{-2} \AA^{-1}}$ and 
a mean of $(1.7\pm0.4)\times10^{-14}~{\rm erg~s^{-1} cm^{-2} \AA^{-1}}$.
The NUV flux densities range from $6.8\times10^{-16}$ to
$2.6\times10^{-13}~{\rm erg~s^{-1} cm^{-2} \AA^{-1}}$ with a median of 
$5.1\times10^{-15}~{\rm erg~s^{-1} cm^{-2} \AA^{-1}}$ and a 
mean of $(1.3\pm0.3)\times10^{-14}~{\rm erg~s^{-1} cm^{-2} \AA^{-1}}$.  The log 
of the FUV luminosities range from 8.30 
to 10.33, with a median and a mean of $9.45\pm0.04$, where the luminosities are 
expressed in solar units, uncorrected for reddening.  The log of the NUV luminosities 
range from 8.30 to 10.40, with a median of 9.57 and a mean of $9.64\pm0.04$.
For comparison, the characteristic luminosity $L_*$ for the present day FUV luminosity 
function is $10^{9.6}~L_{\odot}$ \citep{wyder05}.  The GOALS sample is thus on 
average only $30\%$ fainter than $L_*$ in the FUV, and the most UV-luminous 
LIRGs in GOALS are ultraviolet luminous galaxies \citep[UVLGs, defined as 
${\rm log(L_{FUV}/L_{\odot})}>10.3$;][]{heckman05}.

The infrared excess IRX is defined as the ratio of IR to FUV flux, most 
commonly expressed in logarithmic units.  When calculating IRX we use $L_{IR}$, 
the total IR luminosity from 8--$1000~\mu$m.  $L_{IR}$ is calculated using IRAS flux 
densities for integrated systems, and is allocated among resolved galaxies using MIPS 
flux density ratios as described above.  IRAS flux densities for GOALS 
systems are taken from \citet{rbgs}, MIPS flux densities for resolved galaxies 
within GOALS systems are taken from \citet{mazz}, and luminosity distances are 
taken from \citet{armus09}.  IRX values range from 1.08 to 3.42, with a median of 2.02 
and a mean of $2.06\pm0.04$.  Derived quantities are presented in Table~3 for 
integrated systems and in Table~4 for resolved galaxies.

The UV continuum slope $\beta$(GALEX) was calculated according to the definition
of \citet{kong}: 
\begin{equation}
\beta(GALEX)=\frac{{\rm log}(f_{\rm FUV})-{\rm log}(f_{\rm NUV})}{-0.182} 
\end{equation}
where $f_{\rm FUV}$ and $f_{\rm NUV}$ are the mean flux densities per unit wavelength.
Values of $\beta$(GALEX) range from -1.28 to 3.5, with a median of -0.16 
and a mean of $0.07\pm0.08$.  Since the GALEX filters have different effective wavelengths 
than previous instruments such as IUE or STIS, the normalization of $\beta$(GALEX) is different 
from previous work (e.g., Meurer et al. 1999; Goldader et al. 2002; see 
Appendix~A for a more detailed discussion and a direct conversion between 
$\beta$(GALEX) and $\beta$(IUE)).  Of the 135 observed systems, 
112 have good quality data ($\sigma_{\beta({\rm GALEX})}<0.5$) and are used 
in the subsequent analysis.  Eleven galaxies in the 
\citet{meurer} sample are included in GDP.  These systems allow us 
to recreate the linear portion of the IRX-$\beta$(GALEX) relation for (sub-LIRG) starburst 
galaxies, hereafter referred to as the starburst relation.  The eleven \citet{meurer} 
systems in GDP span a range of $-1<\beta({\rm GALEX})<0.5$, and extrapolations 
beyond that range are not necessarily correct.

The IRX-$\beta$(GALEX) plot is shown in Fig.~\ref{irxbeta}.  
As \citet{goldader} discovered, IR-luminous systems tend to lie above the 
starburst relation.  Similarly, as seen 
in \citet{kong}, \citet{cortese06}, and GDP, the starburst relation forms 
an upper envelope for normal galaxies on this plot.  Within the valid range for the 
starburst relation, $15\%$ of LIRGs fall below the relation.  In addition, twelve LIRGs 
with very red UV colors ($\beta>1$) have high IRX values (2.2--3.3) but lie far below a 
linear extrapolation to the starburst relation.  The fit to the 
late-type galaxy sample of \citet{cortese06} provides a particularly clean
separation between (U)LIRGs and sub-LIRGs in Fig.~\ref{irxbeta}.  The 
shallower slope better matches the distributions of GOALS subpopulations 
($L_{IR}<10^{11.4}$, $10^{11.4}<L_{IR}<10^{11.8}$, and $L_{IR}>10^{11.8}$),
with the more luminous subpopulations having 
larger separations in IRX from the Cortese relation.  The best fit slope 
for the GOALS data is ${\rm IRX} = (0.46\pm0.06)\beta+(2.1\pm0.1)$, shallower than the 
Cortese relation by 0.24 but offset to higher IRX by 0.8 at $\beta = 0$.

\subsection{Star Formation Rates}

The combination of IRAS $L_{IR}$ and GALEX FUV measurements provide an 
estimate of the total (obscured plus unobscured) star formation rate 
\citep[SFR;][]{kennicutt98,dale07}.    The contribution of the FUV 
to the measured SFR ranges from 0.2\% to 17.9\%, with a median of 2.8\% 
and a mean of $4.0\pm0.4$\%.   A histogram showing the ratio of UV-derived SFR to that 
derived from the combination of UV and IR luminosity is shown in 
Fig.~\ref{uvsfr}a.  Calculations relating to SFR do not include galaxies with IRAC 
colors that are consistent with the presence of
a strong AGN \citep{stern05}.  The distribution of the FUV contribution to SFR is consistent with 
previous work \citep*{ss00,sse00}.  The FUV contribution to SFR is small for LIRGs 
and ULIRGs, and decreases as $L_{IR}$ increases (Fig.~\ref{uvsfr}b).  The
Spearman rank correlation coefficient $r_s=-0.47$, with significance
$3.6\times10^{-6}$ indicating a significant correlation, although the
relation is clearly non-linear.  Galaxies with larger infrared  
luminosity have a higher fraction of their measured star-formation emerging
in the far-infrared, with a corresponding lower fraction emerging in
the far-ultraviolet.  As a function of IR luminosity, the median (mean) contribution of 
the FUV to the measured SFR is 3.3\% (4.6\%) for systems with $L_{IR}<10^{11.8}$,
and drops to 1.9\% (2.0\%) for systems with $L_{IR}>10^{11.8}$.

IRAC $3.6~\mu$m and 2MASS K band photometry were used to 
estimate the stellar mass of each galaxy \citep{lacey08}.  The mass estimates 
derived from K band data were used where possible.  For the galaxies without reliable
K-band photometry, 
the masses estimated from $3.6~\mu$m data were scaled by the median ratio of 
mass(K)/mass(3.6) from galaxies measured at both wavelengths.  Stellar masses range from
$4.3\times10^{10}$ to $6.4\times10^{11}~M_{\odot}$, with a median of 
$1.4\times10^{11}~M_{\odot}$ and a mean of $(1.63\pm0.09)\times10^{11}~M_{\odot}$.
The specific star formation rate (SFR per unit mass; 
SSFR) ranges from $5.5\times10^{-11}$ to $3.5\times10^{-9}~{\rm yr}^{-1}$, 
with a median of $3.9\times10^{-10}~{\rm yr}^{-1}$ and a mean of 
$(6.2\pm0.7)\times10^{-10}~{\rm yr}^{-1}$.  These correspond to mass doubling timescales 
of 18~Gyr to 290~Myr, with a median of 2.6~Gyr.

The Spitzer Infrared Nearby Galaxy Survey 
\citep[SINGS;][]{sings} provides a useful comparison sample of lower luminosity galaxies 
observed with both GALEX and Spitzer.  The upper bound of the SSFRs measured 
for SINGS galaxies is approximately $3\times10^{-10}~{\rm yr}^{-1}$ \citep{dale07}.  
The IR/UV ratio, a useful observational measure of dust extinction, is defined as 
\begin{equation}
{\rm IR/UV} = \frac{L_{IR}}{\nu{L}_{\nu}(FUV)+\nu{L}_{\nu}(NUV)}
\end{equation}
and ranges from 5.8 to 813, with a median of 39.
Figure~\ref{iruvssfr} compares the IR/UV ratio against SSFR for both
the GOALS and SINGS samples \citep{dale07}.  In GOALS systems, the IR/UV ratio 
is correlated with SSFR ($r_s=0.55$, significance $2\times10^{-8}$), with large 
scatter: LIRGs and ULIRGs with high SSFR also have high IR/UV ratios.  The two 
quantities are anti-correlated ($r_s=-0.61$, significance $1\times10^{-6}$) for SINGS 
galaxies with ${\rm SSFR}>10^{-11}~{\rm yr}^{-1}$.  A handful of SINGS galaxies 
have IR/UV ratios which are as high as seen in the GOALS sample, but their SSFRs 
are significantly lower.  Taken together, the GOALS and SINGS sources span nearly 
four orders of magnitude in IR/UV at high SSFR ($>10^{-10}~{\rm yr}^{-1}$), probing 
very different star forming populations.

To investigate subpopulations of the GOALS sample in SSFR, we define bins 
with ${\rm SSFR}<3\times10^{-10}~{\rm yr}^{-1}$, 
$3\times10^{-10}<{\rm SSFR}<6\times10^{-10}~{\rm yr}^{-1}$,
and ${\rm SSFR}>6\times10^{-10}~{\rm yr}^{-1}$: galaxies which span the same range of 
SSFR as the SINGS sample, galaxies with up to twice the SSFR as the most 
extreme SINGS galaxies, and galaxies with more than twice the SSFR of the 
most extreme SINGS galaxies, respectively.  These subpopulations are 
plotted on the IRX-$\beta$(GALEX) diagram in Fig.~\ref{ssfr}.
The systems with higher SSFR have higher median offsets from the 
starburst relation than systems with lower SSFR.  Median $\beta$ values are $-0.2\pm0.2$, 
$0.1\pm0.1$, and $-0.20\pm0.09$ (high, medium, and low SSFR bins, respectively).  
Median IRX values are $2.35\pm0.09$, $2.10\pm0.08$, and $1.81\pm0.06$, respectively.  
Systems with $\beta<0.5$ allowing a direct comparison to the starburst relation have median 
vertical deviations of $0.9\pm0.1$, $0.4\pm0.1$, and $0.37\pm0.09$ respectively.

\subsection{Resolved Systems}

A number of the interacting LIRGs in GOALS are near enough to resolve with both 
GALEX and Spitzer and derive IR and UV properties for each galaxy.
Derived quantities for the galaxies in resolved systems are presented 
in Table~4.  The component galaxies of resolved pair/triple systems are 
plotted on the IRX-$\beta$(GALEX) diagram in 
Fig.~\ref{split}.  Many GOALS systems consist of a LIRG with one or more sub-LIRG 
companions.  The sub-LIRG galaxies are on average 
consistent with the GDP sample.  LIRGs are on average offset above the starburst relation,
with $L_{IR}>10^{11.4} L_{\odot}$ systems having larger offsets than lower luminosity 
LIRGs.  For systems with $\beta<0.5$, median offsets are $1.1\pm0.2$ and $0.4\pm0.1$ 
for the $L_{IR}>10^{11.4} L_{\odot}$ and $10^{11}<L_{IR}<10^{11.4} L_{\odot}$ 
populations, respectively.  An individual galaxy in general does not lie in the same region 
of the IRX-$\beta$ diagram as the LIRG system of which it is a component; see \S4.3 for 
further discussion.

%
Of the 18 resolved systems for which masses could be estimated, the 
median mass ratio of the galaxy companions is 2.6:1, with a range from 1.1:1 to 8.1:1.  
The high mass
component of these pairs/triples is, on average, offset above the
starburst relation (Fig.~\ref{mass}), while the low mass components are, on 
average, consistent with the starburst relation.  For systems with $\beta<0.5$, 
median offsets are $0.9\pm0.1$ and $0.3\pm0.1$ for the high mass and low mass 
components, respectively.

\section{Discussion}

The complete sample of the nearest LIRGs and ULIRGs that comprise GOALS is ideal 
for studying the relationship between the IR and UV properties of 
luminous infrared galaxies.
A key diagnostic tool which we explore in this paper is the IRX-$\beta$(GALEX) 
diagram, comparing the IR excess (ratio of IR to FUV emission) to the 
FUV-NUV color parameterized as the power-law slope $\beta$(GALEX).  If a 
class of galaxies, such as starburst 
galaxies, follows tight relations on this diagram, then the measurement 
of the rest-frame UV color allows IRX and thus ${\rm L}_{\rm IR}$ to 
be derived.  This is of particular interest at high redshift, where ${\rm L}_{\rm IR}$
can only be directly measured using far-infrared and submillimeter observations 
but rest-frame UV observations can be made at visual wavelengths in deep surveys.  
Since LIRGs contribute significantly to the star-formation activity at high 
redshift \citep[e.g.,][]{magnelli09}, understanding the IRX-$\beta$(GALEX)
relation in this population is extremely important.  The IRX-$\beta$(GALEX) diagram, and the 
combination of UV and IR data more generally, provide an indication 
of the obscuration to the young stars (or active nucleus) within a galaxy.  This can provide 
a rough test of the evolutionary sequence in which some starburst galaxies 
transition from LIRGs to ULIRGs to QSO hosts over the course of a major merger 
event as the dust and gas is first funneled towards the nuclei fueling a starburst, 
only to be cleared away by the action of AGN and starburst winds in the final 
stages of the transformation to a QSO.

To estimate the importance of high-$\beta$ galaxies among the IR population as
a whole, the fraction of the total IR luminosity integrated over all 629 galaxies in 
the RBGS contributed by the 112 LIRGs and ULIRGs of the GOALS GALEX sample 
is shown as a function of $\beta$ in Fig.~\ref{histbetalir}.  Within the GOALS sample, 
more luminous systems have, on average, larger IRX and redder $\beta$ values 
than less luminous systems while maintaining roughly the same offset from 
the starburst relation.  As shown in 
Fig.~\ref{histbetalir}, a minimum of 19\% of the total infrared luminosity of 
the 629 galaxies that comprise the RBGS is produced in LIRGs and ULIRGs with a $\beta>0$ 
(IUE or GALEX).  These red sources are typically absent from UV-selected 
samples at high redshift, regardless of their estimated IR luminosity.  This is a 
strict lower limit, since there are a number of low-z LIRGs not observed or
detected with GALEX which might have large $\beta$.  

\subsection{Explaining Scatter in the IRX-$\beta$(GALEX) Relation}

The trend for certain populations to have, on average, larger values 
of IRX and redder values of $\beta$(GALEX) (parallel to the starburst relation)
has been explained as a sequence in optical depth \citep{charlot00}.
Thus, on average, more luminous LIRGs and ULIRGs have more extinction than 
less luminous LIRGs, and interacting systems have more extinction than non-interacting 
systems.  This is consistent with the evolutionary scenario mentioned earlier.

We interpret the scatter of LIRGs and ULIRGs in the IRX-$\beta$(GALEX) diagram as follows.  
Deviations to the right of the starburst relation are interpreted as purely the result of 
redder UV colors (extra NUV emission for a given amount of FUV emission), most likely 
due to light from older stellar populations \citep{kong}.  Deviations above the starburst
relation are interpreted as the result of increases in IRX, which we 
define as $\Delta{\rm IRX}$.  This quantity is a measure of the extent 
to which the IR and UV emission become decoupled, for example in heavily 
obscured nuclei which emit strongly in the FIR (UV radiation reprocessed by 
dust) but do not contribute to the observed (escaping)
UV emission.  Like the starburst relation, $\Delta{\rm IRX}$ is not necessarily 
accurate for $\beta({\rm GALEX})<-1$ or $\beta({\rm GALEX})>0.5$.
A minimum of 11\% of the total $L_{IR}$ of the RBGS sample is produced in LIRGs and 
ULIRGs with $\Delta{\rm IRX}>1$, an order of magnitude above the starburst relation.

\citet{cortese06} concluded that attempting to estimate $L_{IR}$ from rest-frame UV 
data of high redshift galaxies will be uncertain by $>50$\% for 
normal galaxies.  We find that using the starburst relation to estimate 
$L_{IR}$ from rest-frame UV observations of LIRGs and ULIRGs would 
on average underestimate $L_{IR}$ by a factor of 2.7 with a range of 
$L_{IR}$(true)/$L_{IR}$(estimated) between 0.2 to 68.  Overestimates 
can be much greater for red UV colors beyond the range of the starburst 
relation ($\beta({\rm GALEX}) > 0.5$), up to a factor of 2400 for a linear extrapolation. 
Previous studies have investigated possible second parameters for 
the scatter of normal galaxies to the right the starburst relation.  Using a sample 
of a wide variety of galaxy types, \citet{seibert05} found no correlation between 
the deviation from the starburst relation and $L_{IR}, L_{UV}, L_{bol}$, or UV and 
optical colors.  Among normal galaxies, any correlation with star formation history is 
weak \citep{kong, cortese06} or nonexistent \citep{seibert05, boquien}.

A number of observables might explain the scatter in $\Delta{\rm IRX}$, providing 
a second parameter to allow more accurate 
measurements of $L_{IR}$ at high redshift as well as physical 
insight into the evolution of LIRGs and ULIRGs.  A central question is what 
mechanism(s) lead to the UV emission being heavily obscured or decoupled 
from the IR emission in (U)LIRGs \citep{goldader} but not in lower luminosity 
starbursts?  Since many LIRGs and essentially all ULIRGs are merger 
remnants with intense, compact, dust-enshrouded nuclear starbursts or AGN, a 
concentration parameter might correlate with IRX-$\beta$(GALEX) scatter 
as an indicator of decoupled IR and UV emission.  Similarly, warm 
IR colors such as IRAS $25~\mu$m/$60~\mu$m might indicate dust in 
close proximity to a powerful UV source (starburst or AGN).  AGN provide another 
possible mechanism to explain scatter from the starburst relation.  The [3.6]-[4.5] 
and [5.8]-[8] IRAC colors \citep{stern05} can be used as an indicator of AGN emission.  
Systems identified as potential AGN might correlate with larger IRX above 
what the starburst relation would predict.
Finally, although heightened IRX in a population of LIRGs and ULIRGs is most 
logically explained by elevated IR emission, it is possible for low 
UV emission to produce the same result.

As shown in Fig.~\ref{corr},
$\Delta{\rm IRX}$ increases with IR luminosity for 
$L_{IR}\gtrsim10^{10}~L_{\odot}$, with considerable scatter.  GOALS systems with 
IRAC colors that may indicate the presence of an AGN tend to have larger IRX ratios by 
a factor of six.
No correlation is found 
between $\Delta{\rm IRX}$ and any of the following 
quantities: IRAS $25~\mu$m/$60~\mu$m color, IRAS $60~\mu$m/$100~\mu$m color, 
Spitzer $8~\mu$m/$24~\mu$m color, ${\rm L_{FUV}}$, $8~\mu$m concentration (1~kpc/Total).  
%
The lack of correlation between $\Delta{\rm IRX}$ and global parameters other 
than $L_{IR}$ suggest that the decoupling between UV and 
IR emission takes place on sub-kpc scales in most LIRGs and ULIRGs, well below our 
resolution with GALEX and Spitzer MIPS~$24~\mu$m, which is 2.6~kpc 
($6^{\prime\prime}$) at the median distance of the GOALS sample (89~Mpc).  
Future studies (e.g. with Herschel and HST) at higher spatial resolution in the FIR 
and UV will be able to investigate this further.  Such studies have already been done 
for a few nearby quiescent star-forming galaxies.  \citet{boquien} found that variation in 
dust extinction curves and geometry is the most important factor determining the location 
of individual star-forming regions on the IRX-$\beta$ diagram.  \citet{munozmateos} 
examined radial profiles of all available SINGS galaxies and found that star formation 
history is the primary driver determining the position on the IRX-$\beta$ 
diagram of a radial annulus within a galaxy.  The lack of correlation 
between $\Delta{\rm IRX}$ and FIR colors suggests that when dust is 
close to the heating source (producing warm FIR colors), that source is 
obscured and the UV color $\beta$(GALEX) increases along with IRX.
Galaxies with positive $\Delta{\rm IRX}$ span a range of ${\rm log(L_{FUV})}$ 
from 8.6 to 10.3 uniformly.  The range of FUV luminosities 
indicates that LIRGs and ULIRGs with large $\Delta{\rm IRX}$ value are IR-bright, 
not UV-faint.  Figure~\ref{l1600} shows the IR/UV ratio and $\Delta{\rm IRX}$ 
plotted against the $1600{\rm \AA}$ luminosity (derived by linear interpolation between 
FUV and NUV).

In order to explore the dependence of IRX and $\beta$ on the morphological 
properties of LIRGs, all GOALS systems were visually classified as either interacting or 
non-interacting based on the inspection of the Spitzer IRAC 3.6um images.  
A galaxy was deemed interacting if it exhibited a tidal bridge or tail, double nuclei,  
multiple galaxies in a common envelope or a disturbed morphology.
The interacting and non-interacting subpopulations 
are shown on the IRX-$\beta$(GALEX) diagram in Fig.~\ref{merger}.  Although
the median position of the interacting population has redder $\beta$ (median 0.0 vs. 
$-0.19$) and larger IRX (2.01 vs. 1.86) than the non-interacting population, the 
two populations are consistent with being drawn from the same distribution.  
The galaxies with the lowest IRX are predominantly interacting, and these systems 
are among the most UV-luminous sources in the GOALS sample with 
${\rm log(L_{FUV}/L_{\odot})}\gtrsim10$.

UVLGs are an interesting type of object to compare with (U)LIRGs since they are objects 
with extremely high SFR but little dust obscuration.
Five LIRGs in our sample are also UVLGs or near-UVLGs 
(${\rm L_{FUV}}\geq 10^{10.2} L_{\odot}$): Arp~256, VV~114, Arp~240, 
NGC~6090, and CGCG~448-020.  The stellar masses of these systems range from 
$11.1\leq {\rm log(M_{stellar}/M_{\odot})}\leq11.5$.  SFR derived from the combination 
of UV and IR luminosities range from $1.8\leq{\rm log(\frac{SFR}{M_{\odot} yr^{-1}})}\leq2.2$, 
and SSFR range from $-9.6\leq{\rm log(SSFR/yr^{-1})}\leq -8.9$.  The sample of \citet{heckman05} 
is divided into Large UVLGs and Compact UVLGs, which have mass ranges of 
$10.5\leq {\rm log(M_{stellar}/M_{\odot})}\leq11.1$ and 
$9.5\leq {\rm log(M_{stellar}/M_{\odot})}\leq10.7$ respectively, SFR ranges of 
$0.6\leq{\rm log(\frac{SFR}{M_{\odot} yr^{-1}})}\leq1.2$ and 
$0.6\leq{\rm log(\frac{SFR}{M_{\odot} yr^{-1}})}\leq1.4$ respectively, and SSFR ranges of
$-10.5\leq{\rm log(SSFR/yr^{-1})}\leq -9.5$ and
$-9.8\leq{\rm log(SSFR/yr^{-1})}\leq -8.6$ respectively.  The LIRG UVLGs have larger 
stellar masses and considerably higher SFR than either the Large or Compact UVLG 
samples as a whole.  The LIRG UVLGs have similar SSFR to the Compact 
UVLG sample, the latter group being considered as local 
analogs to high-redshift Lyman Break Galaxies \citep[LBGs; see][]{overzier09}.  

\subsection{Optical and UV-selected (U)LIRG Samples}

Figures~\ref{iruvssfr} and \ref{ssfr} show that, on average, (U)LIRGs with 
high SSFR have larger IRX and IR/UV and redder $\beta$ than (U)LIRGs 
with lower SSFR, implying greater extinction by dust in the high SSFR systems.  
The GOALS sample spans the same range of SSFR as the UV-selected 
sample of \citet{buat09}.  However the UV-selected sample does not include 
galaxies with high IRX (${\rm log(IRX)}\gtrsim2.0$), which comprise $48\%$ 
of the GOALS sample.  The LIRGs in the LBG sample of \citet{buat09} include 
some systems similar to the GOALS UVLGs, while the majority have higher
$L_{UV}$ and low IRX.

The inverse of SSFR provides a doubling timescale for the stellar mass 
of a galaxy.  The range for GOALS systems (excluding those with IRAC colors 
suggesting a possible AGN) is from 18~Gyr 
to 290~Myr, with a median of 2.6~Gyr.  \citet{kaviraj} fit double-burst 
star formation history models to a large sample of SDSS-selected LIRGs 
out to $z=0.2$, finding average burst ages of 7~Gyr and 1~Gyr.  The 
43 systems in common between \citet{kaviraj} and the GOALS GALEX sample 
are consistent with being drawn from the same distribution in $\beta$ and IRX as 
the entire GOALS GALEX sample.

\subsection{Resolved Systems and Implications for Unresolved LIRGs at High Redshift}

As emphasized by \citet*{char04}, individual galaxies in interacting 
systems can have very different far-infrared and UV properties leading
to incorrect assumptions about the system as a whole when viewed as a 
single unresolved system (e.g. at high redshift).  In particular, these authors note that the 
mid-IR/UV ratios of the components of the Arp~299 and VV~114 
systems vary by well over an order of magnitude between the 
individual interacting galaxies.  Our combined GALEX and Spitzer
observations of the GOALS sample shows that this situation exists in a 
significant number of LIRG systems at low redshift.  We define a source that 
produces at least twice as much luminosity as the companion to be dominant 
at that wavelength.  Among LIRGs which can be resolved into interacting galaxies, 
approximately $32\%$ consist of one galaxy which dominates the IR 
luminosity while a companion dominates the UV (hereafter referred to as 
``VV~114-like" systems).  Extrapolating to number counts at $z\geq1$ as in 
\citet{char04}, this implies that as many as $15-30$\% of high redshift galaxies 
are unresolved VV~114-like systems.  

In $21\%$ of resolved systems, a single galaxy dominates both 
the IR and UV emission (such as Arp~182, for example). 
On average the $\Delta$IRX value of the dominant 
galaxy is over four times larger than that of IR-dominant galaxies of similar 
UV color in a VV~114-like system.  If we look at the masses of resolved pairs, 
the $\Delta$IRX of the more massive galaxy 
is on average four times greater than that of the less massive galaxy.  These 
are independent effects: the IR dominant galaxy in a resolved system is likely
to dominate the mass of the system regardless of its contribution to the UV 
luminosity of the system.  If we 
make the simplistic assumption that LIRG mergers form a single 
evolutionary sequence, our observations suggest that the phase in which 
the component galaxies have comparable IR and UV emission is 50\% longer 
than the phase in which a single galaxy dominates both wavelengths.
Furthermore, the fact that the high mass component is above the  
starburst relation would also be consistent with the fact that a synchronization of the  
nuclear starbursts in the two interacting galaxies is rare. 

The ability to visualize merger simulations at observed 
wavelengths from the FUV to the FIR will facilitate the 
interpretation of data sets such as that presented in this 
paper.  The SUNRISE code of \citet{jonsson06} may help 
answer outstanding questions such as: What types of 
mergers (and what fraction of viewing orientations) consist 
of an IR-dominant LIRG with a UV-dominant companion?  What 
mergers consist of a LIRG which dominates both IR and UV 
relative to its companion?  How long do these phases last?
Do certain types of progenitor galaxies (Hubble type, mass 
ratio, gas fraction, orbit, etc.) lead to different 
observables (IRX, $\beta$(GALEX), IR or UV fraction, SSFR, etc.) 
during the merger?  

Although the different definitions of $\beta$(GALEX) 
preclude a direct comparison, the GOALS sample appears to be 
generally consistent with the merger simulations shown in 
\citet{jonsson06}.  In particular, the ULIRG simulations predict an IRX that is up to a factor 
of ten times greater than starburst galaxies with a narrow range of blue to intermediate 
UV colors.  The GOALS ULIRGs within the same range of $\beta$(IUE) have a 
median $\Delta$IRX of 0.9.

\section{Conclusions}

We present a comparison of the UV and infrared properties of 135 LIRGs and ULIRGs
in the GOALS sample observed by GALEX and Spitzer.  We find that:
\begin{itemize}
\item{LIRGs have larger IR excesses than lower luminosity 
galaxies of similar UV color.  On average, more luminous LIRGs and ULIRGs have 
larger IRX and redder colors.} 
\item{The contribution of the FUV to the measured SFR is on average $4\%$; 
UV emission alone is not a reliable indicator of the SFR for LIRGs.}
\item{The median SSFR of the GOALS sample ($3.9\times10^{-10}~{\rm yr}^{-1}$, 
corresponding to a mass doubling timescale of 2.6~Gyr) is 
approximately equal to the maximum SSFR seen in lower luminosity galaxies,
however the median IR/UV ratio (39) for GOALS galaxies is more than an order of 
magnitude greater.}
\item{Deviations from the starburst IRX-$\beta$(GALEX) relation $\Delta$IRX 
increase with IR luminosity for $L_{IR}\gtrsim10^{10}~L_{\odot}$, with considerable 
scatter.  LIRG systems with IRAC colors that may indicate the presence of an AGN 
have average IRX ratios a factor of six larger than the rest of the sample.  $\Delta$IRX
is not strongly correlated with IRAS $25~\mu$m/$60~\mu$m color, IRAS 
$60~\mu$m/$100~\mu$m color, Spitzer $8~\mu$m/$24~\mu$m color, $L_{FUV}$, or 
$8~\mu$m concentration (1~kpc/Total).}
\item{A minimum of 19\% of the total $L_{IR}$ of the RBGS sample is produced in LIRGs 
and ULIRGs with $\beta>0$, sources that are typically absent from UV-selected samples at 
high redshift.  A minimum of 11\% of the total $L_{IR}$ of the RBGS sample is produced in LIRGs 
and ULIRGs with $\Delta{\rm IRX}>1$, an order of magnitude above the starburst relation.}
\item{Using the starburst IRS-$\beta$ relation to estimate $L_{IR}$ from rest-frame UV 
observations of LIRGs and ULIRGs would underestimate $L_{IR}$ by a factor of three 
on average, with a wide range (factors of 0.2--68) of possible under- or over estimates, 
particularly for red UV colors (large values of $\beta$) where $L_{IR}$ could be overestimated 
by as much as a factor of 2400 using a linear extrapolation of the starburst relation.}
\item{The UV and IR properties of GOALS systems are qualitatively consistent with 
an evolutionary picture in which some galaxies transition from LIRGs to 
ULIRGs over the course of a major merger event.  More luminous galaxies, 
mergers, and galaxies with high SSFR are more heavily obscured than 
less luminous galaxies, non-mergers, and galaxies with lower SSFR.}
\item{Among LIRG systems resolved into individual interacting galaxies, 
pairs in which one galaxy dominates the IR emission while the companion 
dominates UV emission (such as the well-studied LIRG system VV~114) are more common 
than pairs in which one galaxy dominates both wavelengths (32\% and 21\% of the 
sample, respectively).  On average, galaxies which dominate both wavelengths have 
$\Delta$IRX values four times larger than an IR-dominant galaxy in a ``VV~114-like" 
system.  The large fraction of ``VV~114-like" systems has important implications for 
observations of interacting galaxies at high redshift in that the IR and UV properties of 
the unresolved systems can differ by over an order of magnitude from the properties of 
the component galaxies.}
\end{itemize}

\acknowledgments
 
This research has made use of the NASA/IPAC Extragalactic Database (NED)
which is operated by the Jet Propulsion Laboratory, California Institute of
Technology, under contract with the National Aeronautics and Space
Administration.  This research has made use of the NASA/ IPAC Infrared Science 
Archive, which is operated by the Jet Propulsion Laboratory, California Institute 
of Technology, under contract with the National Aeronautics and Space 
Administration.  Based on observations made with the NASA Galaxy Evolution 
Explorer.  GALEX is operated for NASA by the California Institute of 
Technology under NASA contract NAS5-98034.  VC acknowledges partial support 
from the EU ToK grant 39965 and FP7-REGPOT 206469.  We thank Ranga 
Chary, Brian Siana, and Harry Teplitz for helpful discussions.  We thank 
Armando Gil de Paz for making his GALEX background subtraction code 
available, Danny Dale for providing the SINGS data points in Fig.~\ref{iruvssfr}, 
and the anonymous referee for helpful comments.

\appendix
\section{UV Colors}

The UV color of an object can be parameterized in several ways, 
complicating the comparison of results between different data sets.
The UV continuum slope $\beta$ was defined by \citet*{calzetti94} for 
use with IUE spectra.  More recent photometric instruments such as 
STIS \citep{goldader} and GALEX cannot directly measure this spectroscopic 
$\beta$, referred to as $\beta({\rm IUE})$ in the main text of this paper and in 
Fig.~\ref{irxbeta}.  Instead the slope between a NUV data point and a 
FUV data point is measured and labeled $\beta$, referred to as $\beta({\rm 
GALEX})$.  Some authors abandon the UV slope and instead 
measure a conventional color FUV-NUV, expressed in magnitudes 
(e.g., GDP).

Since 11 galaxies from \citet{meurer} are included in GDP, we 
derive an empirical conversion between $\beta({\rm IUE})$ and 
$\beta({\rm GALEX})$:

\begin{equation}
\beta({\rm IUE}) = (-0.3\pm0.1) + (1.6\pm0.2)\beta({\rm GALEX})
\end{equation}
\noindent
This conversion is not necessarily valid outside the range 
$-2<\beta({\rm IUE})<0.5$ or $-1<\beta({\rm GALEX})<0.5$.

%


\begin{deluxetable}{lllllllllll}
\tabletypesize{\tiny}
\tablecolumns{11}
\tablewidth{0pt}
\tablecaption{GOALS GALEX Photometry - Integrated Systems}
\tablehead{
\colhead{System} & \colhead{Alternate Name} & \colhead{RA (J2000)} & \colhead{Dec (J2000)} &
\colhead{${\rm log(L_{IR})}$} & \colhead{$t_{FUV}$} & \colhead{$f_{\nu}$(FUV)} &
\colhead{$\sigma_{FUV}$} & \colhead{$t_{NUV}$} & \colhead{$f_{\nu}$(NUV)} &
\colhead{$\sigma_{NUV}$} \\
\colhead{(1)} & \colhead{(2)} & \colhead{(3)} &
\colhead{(4)} & \colhead{(5)} & \colhead{(6)} &
\colhead{(7)} & \colhead{(8)} & \colhead{(9)} &
\colhead{(10)} & \colhead{(11)}
}
\startdata
NGC 0023	&		&	00h09m53.41s	& $	+25^{\circ}55^{\prime}25.6^{\prime\prime}$	&	11.12	&	3410	&	1.26E-14	&	8.E-17	&	3410	&	1.49E-14	&	3.E-17	\\
NGC 0034	&		&	00h11m06.55s	& $	-12^{\circ}06^{\prime}26.3^{\prime\prime}$	&	11.49	&	119	&	1.53E-14	&	4.E-16	&	119	&	1.05E-14	&	2.E-16	\\
Arp 256	&	MCG-02-01-051-2	&	00h18m50.40s	& $	-10^{\circ}22^{\prime}08.0^{\prime\prime}$	&	11.48	&	1618	&	2.65E-14	&	2.E-16	&	1618	&	2.50E-14	&	6.E-17	\\
ESO 350-IG038	&		&	00h36m52.50s	& $	-33^{\circ}33^{\prime}19.0^{\prime\prime}$	&	11.28	&	3356.05	&	3.95E-14	&	1.E-16	&	3356.05	&	2.31E-14	&	4.E-17	\\
NGC 0232	&		&	00h42m45.82s	& $	-23^{\circ}33^{\prime}40.9^{\prime\prime}$	&	11.44	&	115	&	2.65E-15	&	2.E-16	&	115	&	2.37E-15	&	1.E-16	\\
MCG+12-02-001	&		&	00h54m03.61s	& $	+73^{\circ}05^{\prime}11.8^{\prime\prime}$	&	11.50	&	112	&	3.60E-16	&	3.E-16	&	112	&	1.61E-16	&	1.E-16	\\
NGC 0317B	&		&	00h57m41.67s	& $	+43^{\circ}47^{\prime}33.2^{\prime\prime}$	&	11.19	&	110	&	2.36E-15	&	3.E-16	&	110	&	4.29E-15	&	1.E-16	\\
IC 1623	&	VV 114	&	01h07m47.18s	& $	-17^{\circ}30^{\prime}25.3^{\prime\prime}$	&	11.71	&	1668	&	6.40E-14	&	2.E-16	&	1668	&	4.87E-14	&	8.E-17	\\
\enddata
\tablecomments{Column (1): System name, following the naming convention of \citet{armus09}. 
Column (2): Alternate system name.
Column (3): The best available source right ascension (J2000) in NED as of 
October 2008. Column (4): The best available source declination (J2000) in NED as of 
October 2008. Column (5): The total infrared luminosity in ${\rm log_{10}}$ Solar units 
computed using the IRAS flux densities reported in the RBGS and the luminosity distances 
in \citet{armus09}.  Column (6): GALEX FUV integration time in seconds.  Column (7): GALEX FUV flux density in units of ${\rm erg~s^{-1} cm^{-2} \AA^{-1}}$.  Column (8): GALEX FUV flux density
uncertainty.  Column(9): GALEX NUV integration time in seconds.  Column (10): GALEX NUV flux density in units of ${\rm erg~s^{-1} cm^{-2} \AA^{-1}}$.  Column (11): GALEX NUV flux density
uncertainty.}
\end{deluxetable}

\begin{deluxetable}{lllllllllll}
\tabletypesize{\tiny}
\tablecolumns{9}
\tablewidth{0pt}
\tablecaption{GALEX Photometry - Resolved Components}
\tablehead{
\colhead{System Name} & \colhead{Galaxy Name} & \colhead{RA (J2000)} & 
\colhead{Dec (J2000)} & \colhead{${\rm log(L_{IR})}$} & \colhead{$f_{\nu}$(FUV)} &
\colhead{$\sigma_{FUV}$} & \colhead{$f_{\nu}$(NUV)} &
\colhead{$\sigma_{NUV}$} \\
\colhead{(1)} & \colhead{(2)} & \colhead{(3)} &
\colhead{(4)} & \colhead{(5)} & \colhead{(6)} &
\colhead{(7)} & \colhead{(8)} & \colhead{(9)}
}
\startdata
GOALS J001108.5-120351	&	NGC 0034	&	00h11m06.56s	& $	-12^{\circ}06^{\prime}28.2^{\prime\prime}$	&	11.34	&	1.53E-14	&	4.E-16	&	1.05E-14	&	6.E-16	\\
GOALS J001108.5-120351	&	NGC 0035	&	00h11m10.51s	& $	-12^{\circ}01^{\prime}14.9^{\prime\prime}$	&	10.57	&	6.92E-15	&	3.E-16	&	8.63E-15	&	6.E-16	\\
MCG-02-01-051-2	&	MCG-02-01-052	&	00h18m49.85s	& $	-10^{\circ}21^{\prime}34.0^{\prime\prime}$	&	10.36	&	1.24E-14	&	1.E-16	&	1.05E-14	&	1.E-16	\\
MCG-02-01-051-2	&	MCG-02-01-051	&	00h18m50.90s	& $	-10^{\circ}22^{\prime}36.7^{\prime\prime}$	&	11.45	&	1.42E-14	&	1.E-16	&	1.08E-14	&	1.E-16	\\
NGC 0232	&	NGC 0232	&	00h42m45.83s	& $	-23^{\circ}33^{\prime}41.0^{\prime\prime}$	&	11.37	&	2.65E-15	&	2.E-16	&	2.37E-15	&	5.E-16	\\
NGC 0232	&	NGC 0235A	&	00h42m52.82s	& $	-23^{\circ}32^{\prime}27.8^{\prime\prime}$	&	10.86	&	1.28E-15	&	2.E-16	&	1.19E-15	&	5.E-16	\\
NGC 0317	&	NGC 0317A	&	00h57m39.04s	& $	+43^{\circ}48^{\prime}03.1^{\prime\prime}$	&	8.96	&	2.48E-16	&	3.E-17	&	1.37E-15	&	4.E-17	\\
NGC 0317	&	NGC 0317B	&	00h57m40.41s	& $	+43^{\circ}47^{\prime}32.5^{\prime\prime}$	&	11.19	&	2.12E-15	&	3.E-16	&	2.91E-15	&	9.E-17	\\
\enddata
\tablecomments{Column (1): System name. 
Column (2): Galaxy name.
Column (3): The right ascension (J2000) of the IRAC $8~\mu$m centroid in \citet{mazz}. 
Column (4): The declination (J2000) of the IRAC $8~\mu$m centroid in \citet{mazz}. 
Column (5): The infrared luminosity in ${\rm log_{10}}$ Solar units 
computed using the MIPS flux densities reported in \citet{mazz} and the luminosity distances 
in \citet{armus09}; see text for details.  
Column (6): GALEX FUV flux density in units of ${\rm erg~s^{-1} cm^{-2} \AA^{-1}}$.  
Column (7): GALEX FUV flux density uncertainty.  
Column (8): GALEX NUV flux density in units of ${\rm erg~s^{-1} cm^{-2} \AA^{-1}}$.  
Column (9): GALEX NUV flux density uncertainty.}
\end{deluxetable}

\begin{deluxetable}{llllllllll}
\tabletypesize{\tiny}
\tablecolumns{10}
\tablewidth{0pt}
\tablecaption{Derived Quantities - Integrated Systems}
\tablehead{
\colhead{System} & \colhead{${\rm log(L_{FUV})}$} & \colhead{$\beta$(GALEX)} &
\colhead{$\sigma_{\beta}$(GALEX)} & \colhead{IRX} & \colhead{${\rm \sigma_{IRX}}$} 
& \colhead{Mass} & \colhead{SFR} & \colhead{SSFR} & \colhead{IR/UV} 
\\
\colhead{} & \colhead{} & \colhead{} & \colhead{} & \colhead{} & \colhead{} & 
\colhead{$M_{\odot}$} & \colhead{$M_{\odot} {\rm yr}^{-1}$} & 
\colhead{${\rm yr}^{-1}$} & \colhead{} \\
\colhead{(1)} & \colhead{(2)} & \colhead{(3)} &
\colhead{(4)} & \colhead{(5)} & \colhead{(6)} &
\colhead{(7)} & \colhead{(8)} & \colhead{(9)} &
\colhead{(10)} 
}
\startdata
NGC 0023	&	9.39	&	0.409	&	0.016	&	1.724	&	0.003	&	1.59E+11	&	23.34	&	1.47E-10	&	19.2	\\
NGC 0034	&	9.69	&	-0.907	&	0.078	&	1.799	&	0.013	&	1.10E+11	&	[55.25]	&	[5.04E-10] 	&	31.2	\\
Arp 256	&	10.22	&	-0.139	&	0.016	&	1.262	&	0.003	&	1.80E+11	&	57.11	&	3.18E-10	&	7.6	\\
ESO 350-IG038	&	10.15	&	-1.281	&	0.009	&	1.128	&	0.001	&	4.63E+10	&	[37.02]	&	[7.99E-10]	&	7.2	\\
NGC 0232	&	9.04	&	-0.265	&	0.224	&	2.404	&	0.037	&	1.94E+11	&	48.65	&	2.50E-10	&	108.7	\\
MCG+12-02-001	&	7.90	&	-1.915	&	2.361	&	3.593	&	0.343	&	8.10E+10	&	54.49	&	6.73E-10	&	2348.3	\\
NGC 0317B	&	8.82	&	1.421	&	0.339	&	2.367	&	0.060	&	1.16E+11	&	26.97	&	2.33E-10	&	63.0	\\
IC 1623	&	10.33	&	-0.649	&	0.010	&	1.379	&	0.002	&	1.62E+11	&	[94.09]	&	[5.80E-10]	&	11.2	\\
\enddata
\tablecomments{Column (1): System name. 
Column (2): The total FUV luminosity in ${\rm log_{10}}$ Solar units. 
Column (3): $\beta$(GALEX) calculated as described in \S3.1, Equation~1. 
Column (4): $\beta$(GALEX) uncertainty.  
Column (5): Ratio of IR to FUV flux IRX calculated as described in \S3.1.  
Column (6): IRX uncertainty.  
Column (7): Stellar mass calculated as described in \S3.2.  
Column (8): Star formation rate calculated as described in \S3.2.  Values in brackets should be considered upper limits due to possible AGN contamination.
Column (9): Specific star formation rate calculated as described in \S3.2.  Values in brackets should be considered upper limits due to possible AGN contamination.
Column (10): IR/UV ratio calculated as described in \S3.2, Equation~2. 
}
\end{deluxetable}

\begin{deluxetable}{lllllllll}
\tabletypesize{\tiny}
\tablecolumns{9}
\tablewidth{0pt}
\tablecaption{Derived Quantities - Resolved Components}
\tablehead{
\colhead{Galaxy} & \colhead{${\rm log(L_{FUV})}$} & \colhead{$\beta$(GALEX)} &
\colhead{$\sigma_{\beta}$(GALEX)} & \colhead{IRX} & \colhead{${\rm \sigma_{IRX}}$} &
\colhead{Mass} & \colhead{SFR} & \colhead{SSFR} \\
\colhead{} & \colhead{} & \colhead{} & \colhead{} & \colhead{} & \colhead{} & \colhead{$M_{\odot}$} & \colhead{$M_{\odot} {\rm yr}^{-1}$} & \colhead{${\rm yr}^{-1}$} \\
\colhead{(1)} & \colhead{(2)} & \colhead{(3)} &
\colhead{(4)} & \colhead{(5)} & \colhead{(6)} &
\colhead{(7)} & \colhead{(8)} & \colhead{(9)}
}
\startdata
NGC 0034	&	9.69	&	-0.91	&	0.15	&	1.739	&	0.013	&	8.92E+10	&	[46.91]	&	[5.26E-10]	\\
NGC 0035	&	9.35	&	0.53	&	0.19	&	1.257	&	0.020	&	2.05E+10	&	6.99	&	3.41E-10	\\
MCG-02-01-052	&	9.89	&	-0.39	&	0.03	&	0.468	&	0.004	&	...	&	3.95	&	...	\\
MCG-02-01-051	&	9.95	&	-0.64	&	0.03	&	1.501	&	0.003	&	...	&	48.63	&	...	\\
NGC 0232	&	9.04	&	-0.26	&	0.55	&	2.312	&	0.037	&	1.10E+11	&	39.14	&	3.57E-10	\\
NGC 0235A	&	8.73	&	-0.18	&	1.05	&	1.998	&	0.062	&	8.46E+10	&	[9.21]	&	[1.09E-10]	\\
NGC 0317A	&	7.84	&	4.09	&	0.34	&	1.120	&	0.060	&	8.15E+10	&	0.16	&	1.95E-12	\\
NGC 0317B	&	8.77	&	0.76	&	0.34	&	2.412	&	0.060	&	3.44E+10	&	26.63	&	7.73E-10	\\
\enddata
\tablecomments{Column (1): Galaxy name. 
Column (2): The total FUV luminosity in ${\rm log_{10}}$ Solar units. 
Column (3): $\beta$(GALEX) calculated as described in \S3.1, Equation~1. 
Column (4): $\beta$(GALEX) uncertainty.  
Column (5): Ratio of IR to FUV flux IRX calculated as described in \S3.1.  
Column (6): IRX uncertainty.  
Column (7): Stellar mass calculated as described in \S3.2.  
Column (8): Star formation rate calculated as described in \S3.2.  Values in brackets should be considered upper limits due to possible AGN contamination.
Column (9): Specific star formation rate calculated as described in \S3.2.  Values in brackets should be considered upper limits due to possible AGN contamination.}
\end{deluxetable}

\vbox{
\begin{center}
\includegraphics[width=\textwidth]{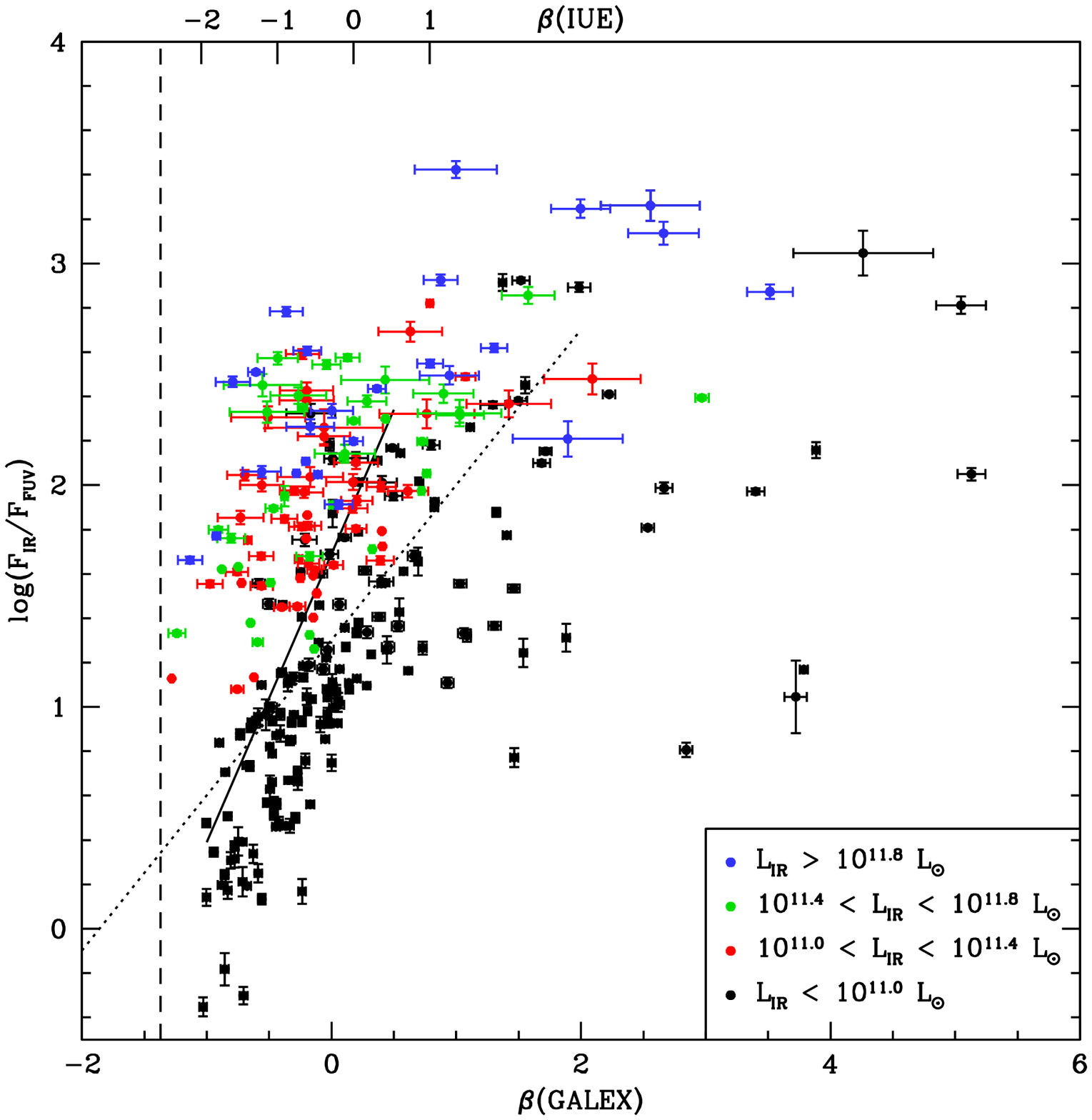}
\figcaption{\small
The IR excess, ${\rm F_{IR}}/{\rm F_{FUV}}$, plotted against the 
UV continuum slope, $\beta$(GALEX).  Black points (from GDP) have 
${\rm L_{IR}}<10^{11} L_{\odot}$, 
red points have $10^{11}<{\rm L_{IR}}<10^{11.4} L_{\odot}$, green points have 
$10^{11.4}<{\rm L_{IR}}<10^{11.8} L_{\odot}$, and blue points have 
${\rm L_{IR}}>10^{11.8} L_{\odot}$.  The solid line is a fit to the starburst 
galaxies of \citet{meurer} which were included in the GDP sample.
The dotted line is the fit to the late-type galaxy sample of 
\citet{cortese06}.  The vertical dashed line is the UV color of a Starburst99 
\citep{sb99} model of a $10^8$ year old starburst population with 
solar metallicity and a Salpeter IMF \citep{salpeter}.
The range of $\beta$(GALEX) in the IUE system of \citet{meurer} is shown at top.
Low and medium luminosity LIRGs (red and green points) fill parameter space 
between normal galaxies and high luminosity LIRGs and ULIRGs (blue points).
\label{irxbeta}
}
\end{center}}

\vbox{
\begin{center}
\includegraphics[width=\textwidth]{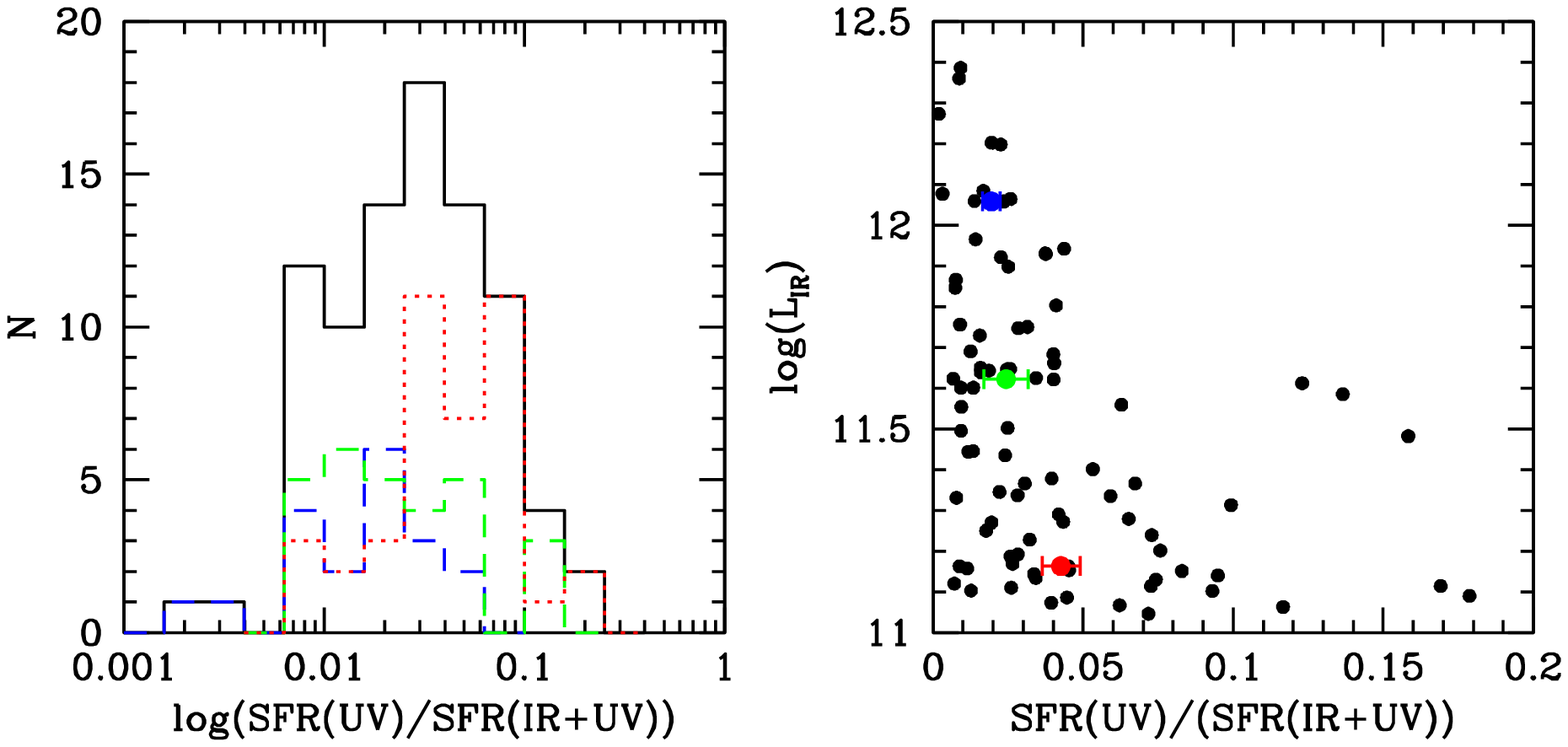}
\figcaption{\small
Left: Histogram showing the ratio of SFR(UV) to SFR(UV+IR).  The solid line is the 
full GOALS GALEX sample.  Colored lines show the GOALS GALEX sample 
divided into luminosity bins as in Fig.~\ref{irxbeta}: $10^{11}<{\rm L_{IR}}<10^{11.4} L_{\odot}$ 
(red dotted line), $10^{11.4}<{\rm L_{IR}}<10^{11.8} L_{\odot}$ (green dashed line), 
and ${\rm L_{IR}}>10^{11.8} L_{\odot}$ (blue dashed line).   The FUV contribution to SFR is 
small for (U)LIRGs and decreases with increasing ${\rm L_{IR}}$.
Right: $L_{IR}$ plotted against the ratio of SFR(UV) to SFR(UV+IR).  Median ratios of the 
star formation rates are shown for each luminosity bin 
(red: $10^{11}<{\rm L_{IR}}<10^{11.4} L_{\odot}$, 
green: $10^{11.4}<{\rm L_{IR}}<10^{11.8} L_{\odot}$, blue: ${\rm L_{IR}}>10^{11.8} L_{\odot}$) 
along with $1\sigma$ standard deviations of the mean.  Although 
anticorrelated (Spearman rank correlation coefficient of -0.47) the correlation is not linear.
\label{uvsfr}
}
\end{center}}

\vbox{
\begin{center}
\includegraphics[width=\textwidth]{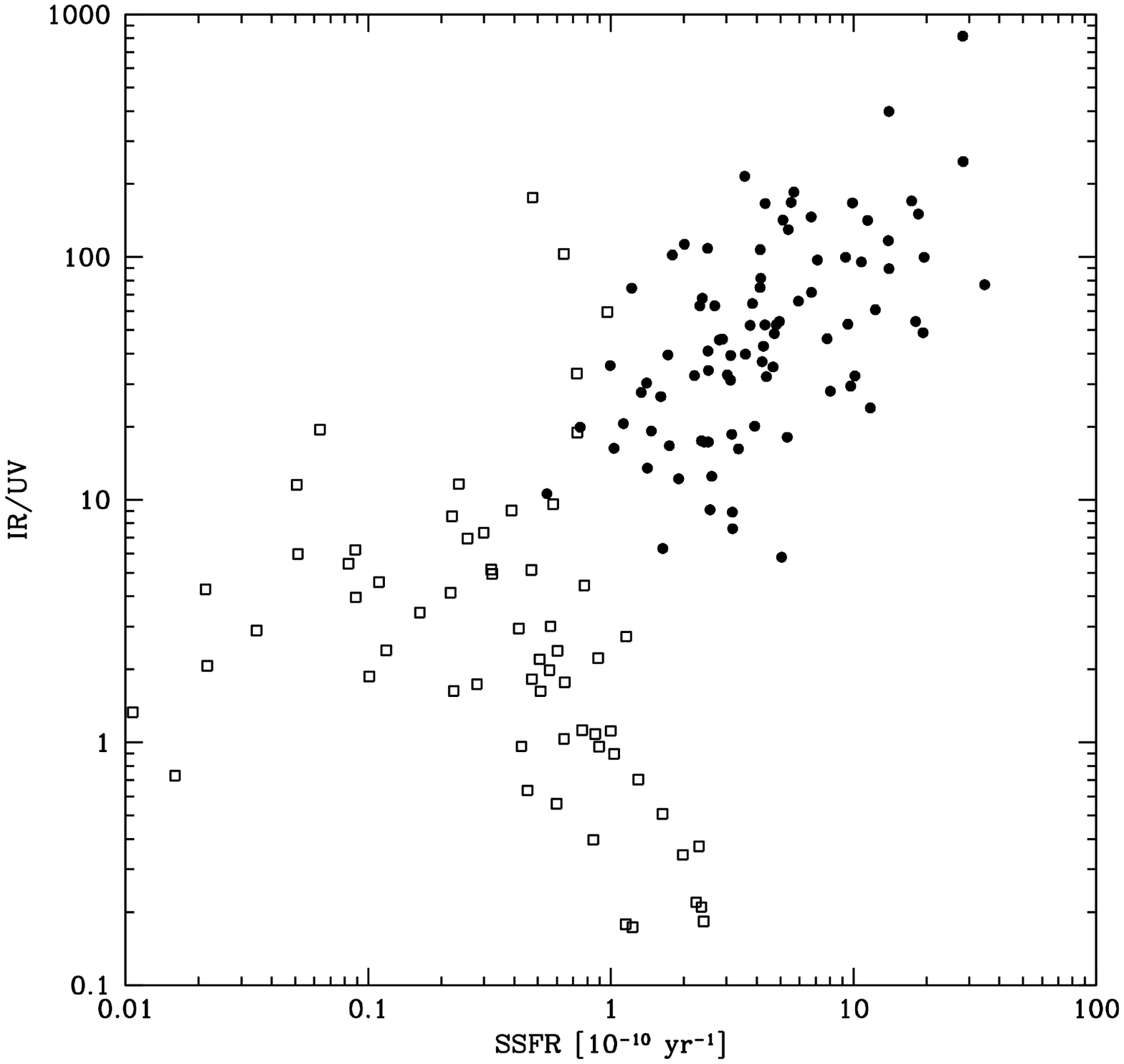}
\figcaption{\small
The IR/UV ratio plotted against specific star formation rate. 
Solid circles are GOALS galaxies (not including those with IRAC 
colors suggesting a strong AGN), while open squares are SINGS galaxies 
\citep{dale07}.  The GOALS outlier at the high IR/UV, high SSFR 
extreme is Arp~220.  LIRGs and ULIRGs have much higher IR/UV ratios 
and SSFR than lower luminosity galaxies, and the two quantities are correlated 
for GOALS systems and anti-correlated for SINGS galaxies with 
${\rm SSFR} > 10^{-11}~{\rm yr}^{-1}$.
\label{iruvssfr}
}
\end{center}}

\vbox{
\begin{center}
\includegraphics[width=\textwidth]{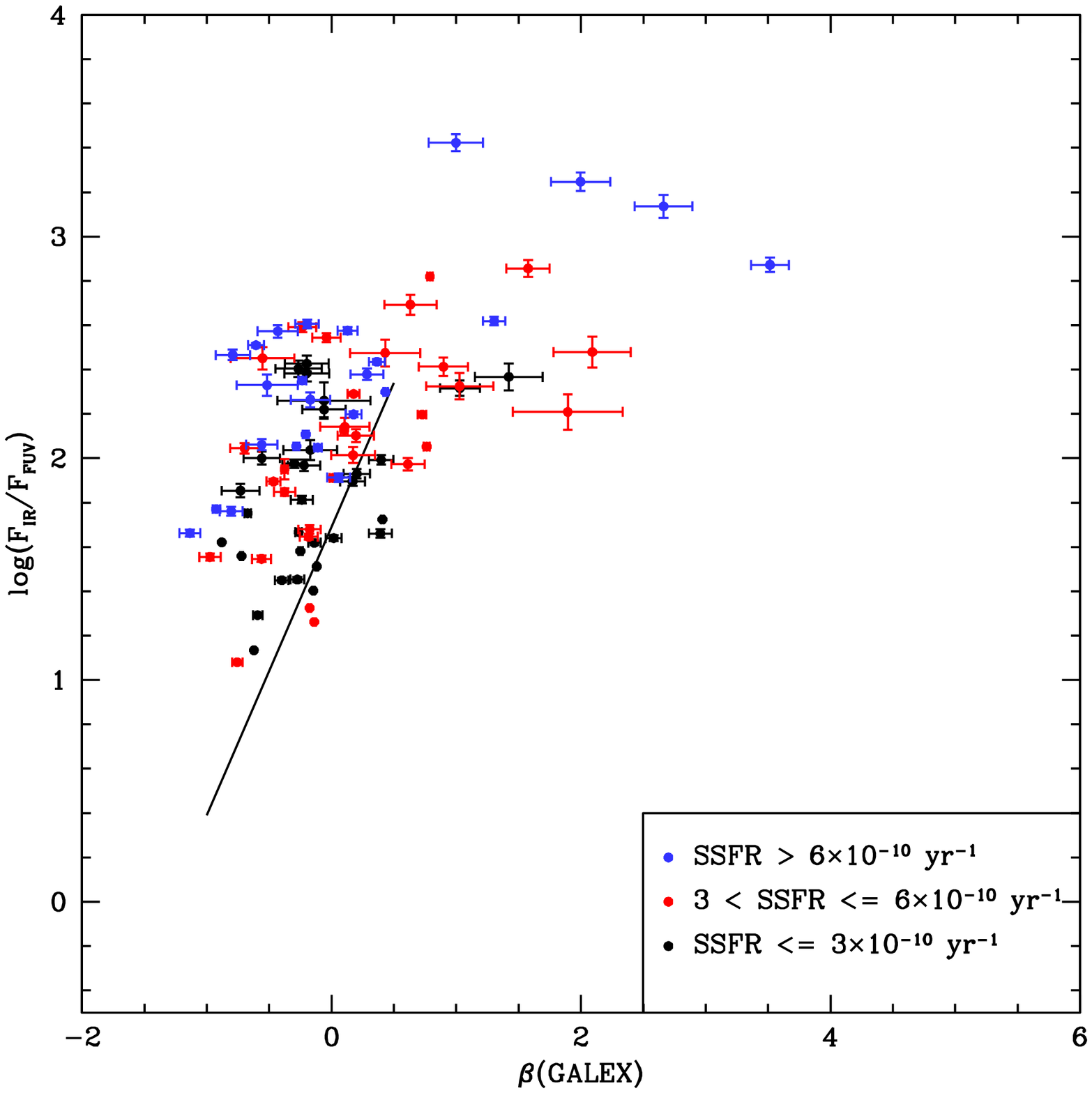}
\figcaption{\small
IRX-$\beta$(GALEX) plot with GOALS systems (not including those with IRAC 
colors suggesting a strong AGN) color-coded by specific SFR: black 
points have SSFR within the range spanned by SINGS galaxies, red points 
have up to twice the SSFR of any SINGS galaxy, and blue points have more 
than twice the SSFR of any SINGS galaxy.  The solid line is the same as in Fig.~\ref{irxbeta}.  
Systems with higher SSFR are systematically redder in $\beta$ and have larger IRX than 
systems with lower SSFR.
\label{ssfr}
}
\end{center}}

\vbox{
\begin{center}
\includegraphics[width=\textwidth]{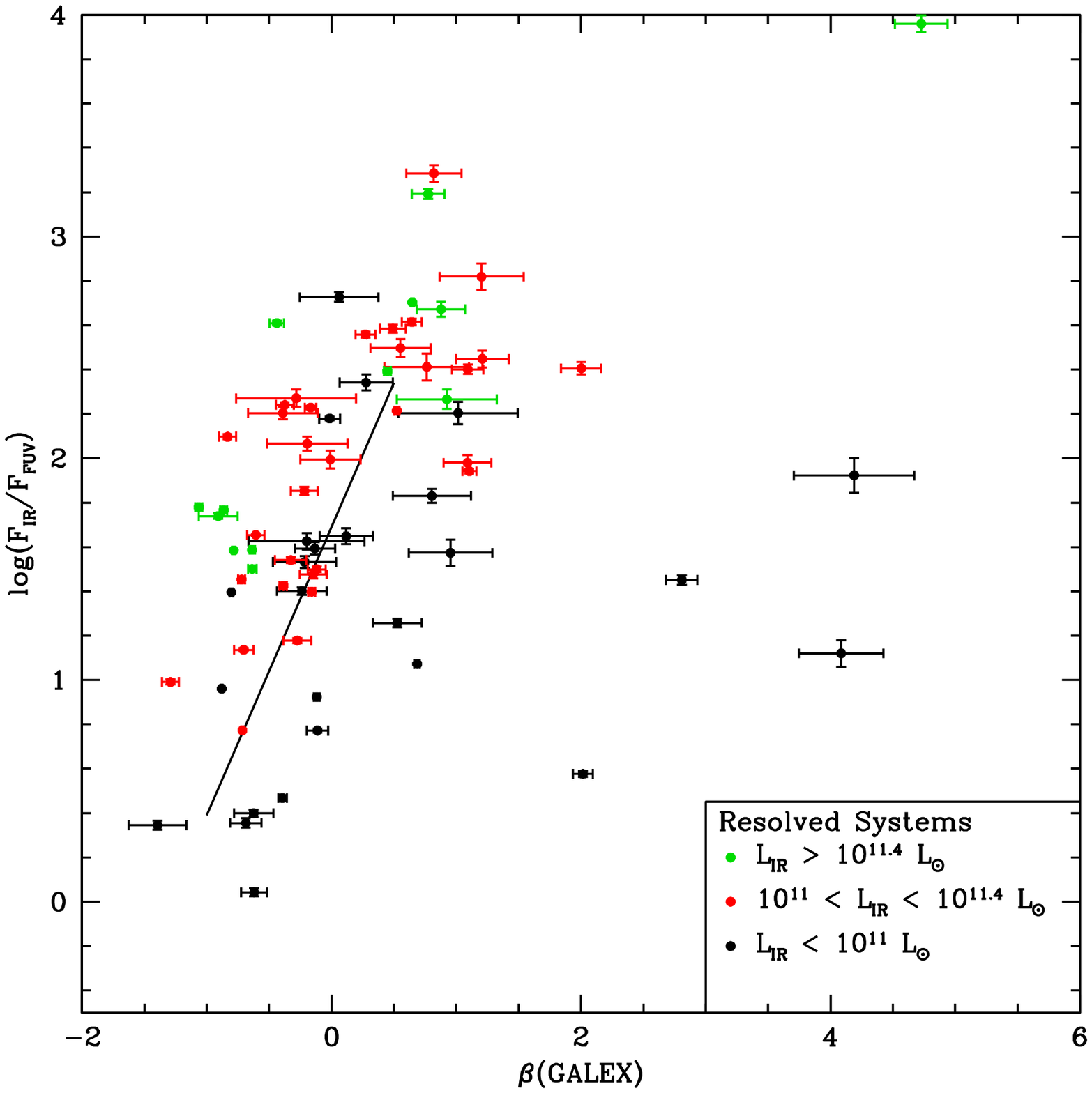}
\figcaption{\small
IRX-$\beta$(GALEX) plot showing the locations of individual galaxies in 
resolved pairs.  As in Fig.~\ref{irxbeta}, black points have 
${\rm L_{IR}}<10^{11} L_{\odot}$, red points have $10^{11}<{\rm L_{IR}}<10^{11.4} L_{\odot}$, 
and green points have $10^{11.4}<{\rm L_{IR}}<10^{11.8} L_{\odot}$.  The solid line 
is the same as in Fig.~\ref{irxbeta}.  Sub-LIRG galaxies are on average consistent with 
the GDP sample.  LIRGs are on average offset above the starburst relation, with 
${\rm L_{IR}}>10^{11.4} L_{\odot}$ systems having larger offsets than lower luminosity LIRGs.
\label{split}
}
\end{center}}

%
\vbox{
\begin{center}
\includegraphics[width=\textwidth]{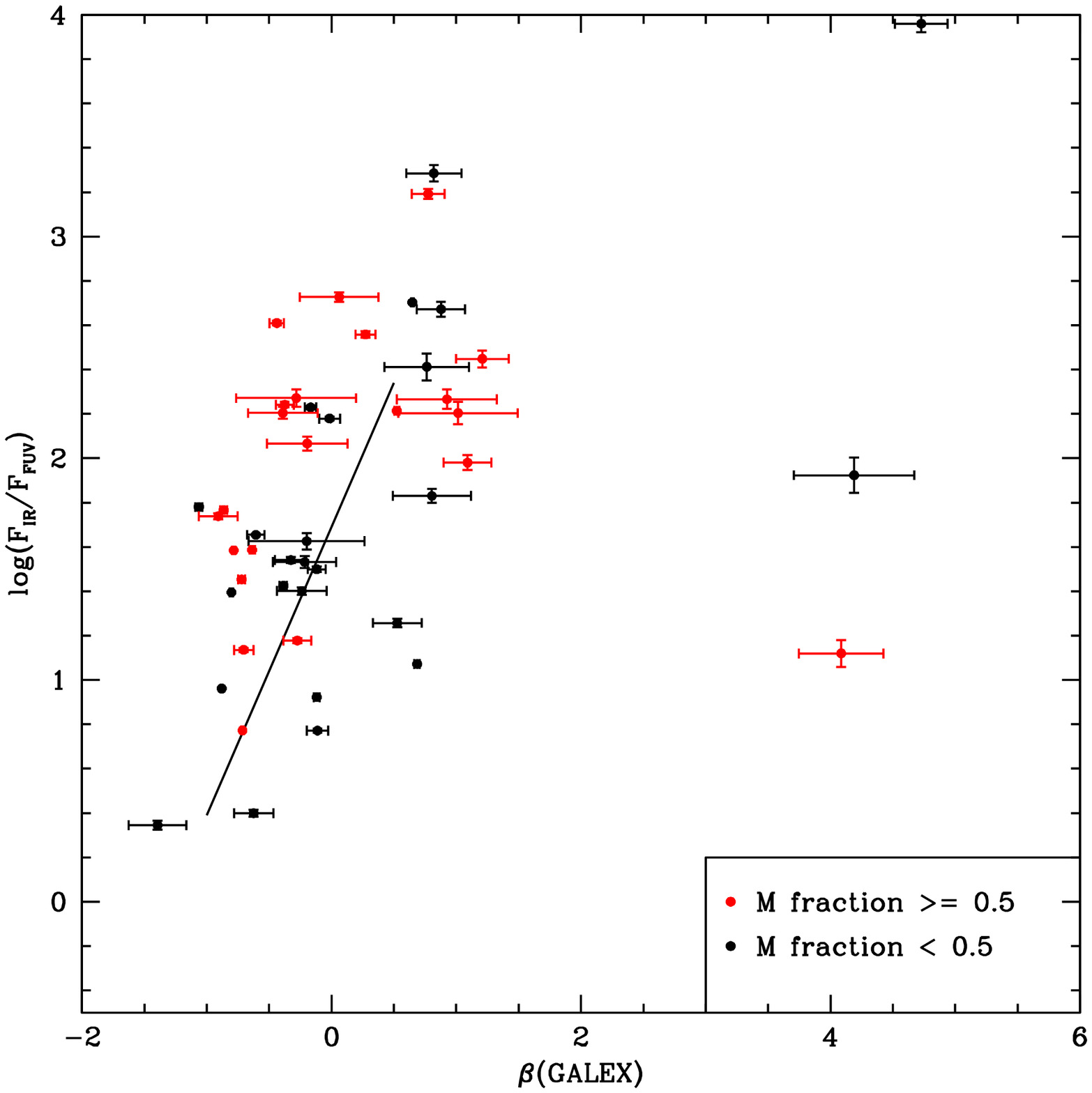}
\figcaption{\small
IRX-$\beta$(GALEX) plot showing the location of individual galaxies in close pairs
for which a mass could be estimated.  The galaxy with $>50\%$ of the mass 
in each system is shown in red, while other galaxies are in black.  
The solid line shows the starburst relation, as in Fig.~\ref{irxbeta}.  On average, the 
high mass galaxy in a system is offset above the starburst relation, while the lower 
mass galaxy lies slightly below the starburst relation.
\label{mass}
}
\end{center}}

\vbox{
\begin{center}
\includegraphics[width=\textwidth]{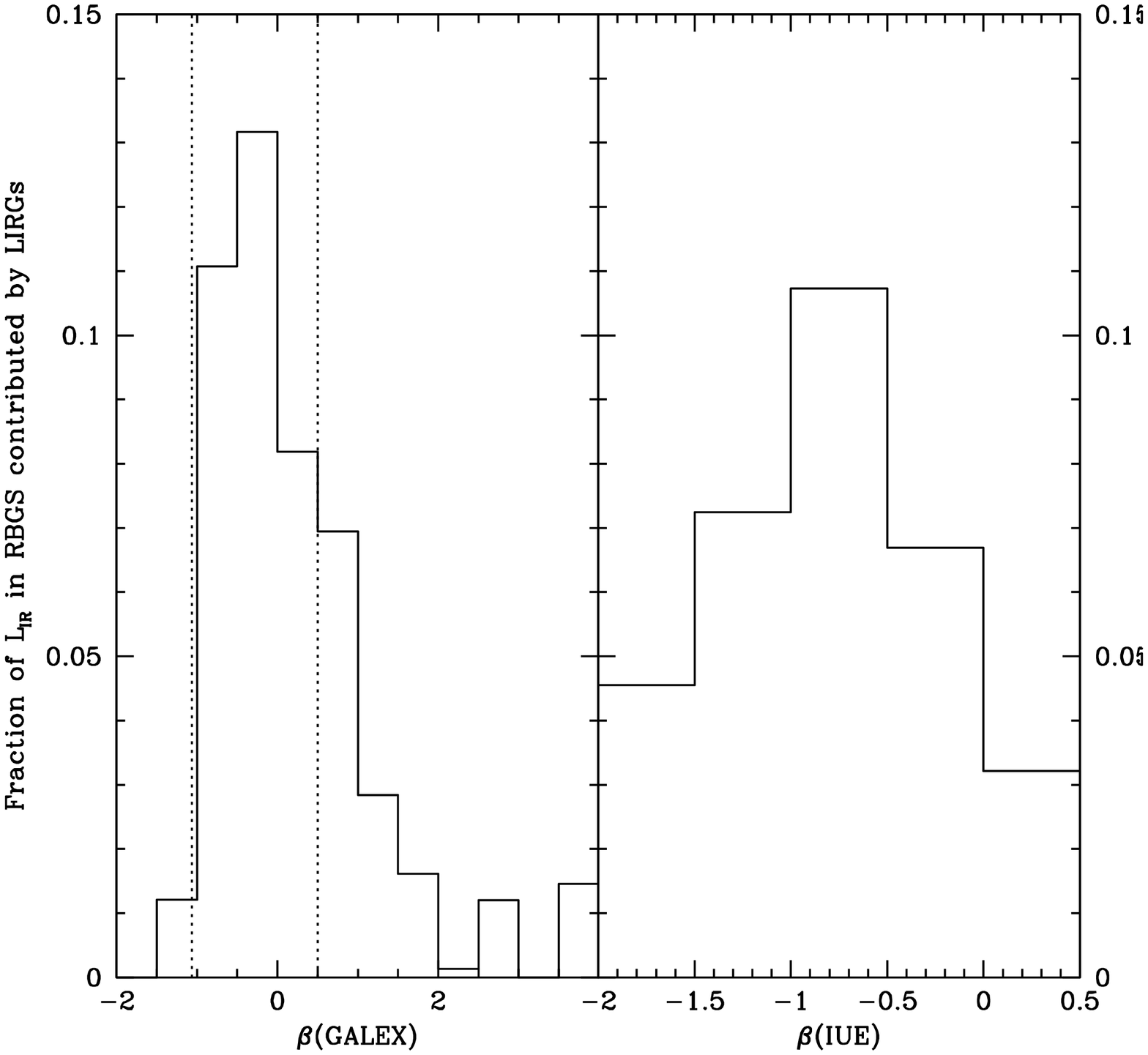}
\figcaption{\small
The fraction of total IR luminosity summed over all 629 systems in the RBGS sample 
contributed by LIRGs 
and ULIRGs with known UV colors (the GOALS GALEX sample, 112 systems).  The IR 
luminosity fraction defined in this way is shown as a function of $\beta$(GALEX) (left 
panel) and $\beta$(IUE) (right panel) over the range of the conversion given in 
Appendix~A.  The dotted lines in the left panel mark the range of $\beta$(GALEX) 
shown in the right panel.  At least $19\%$ of the IR luminosity of the RBGS is produced 
by (U)LIRGs with red UV colors ($\beta>0$).
\label{histbetalir}
}
\end{center}}

\vbox{
\begin{center}
\includegraphics[width=\textwidth]{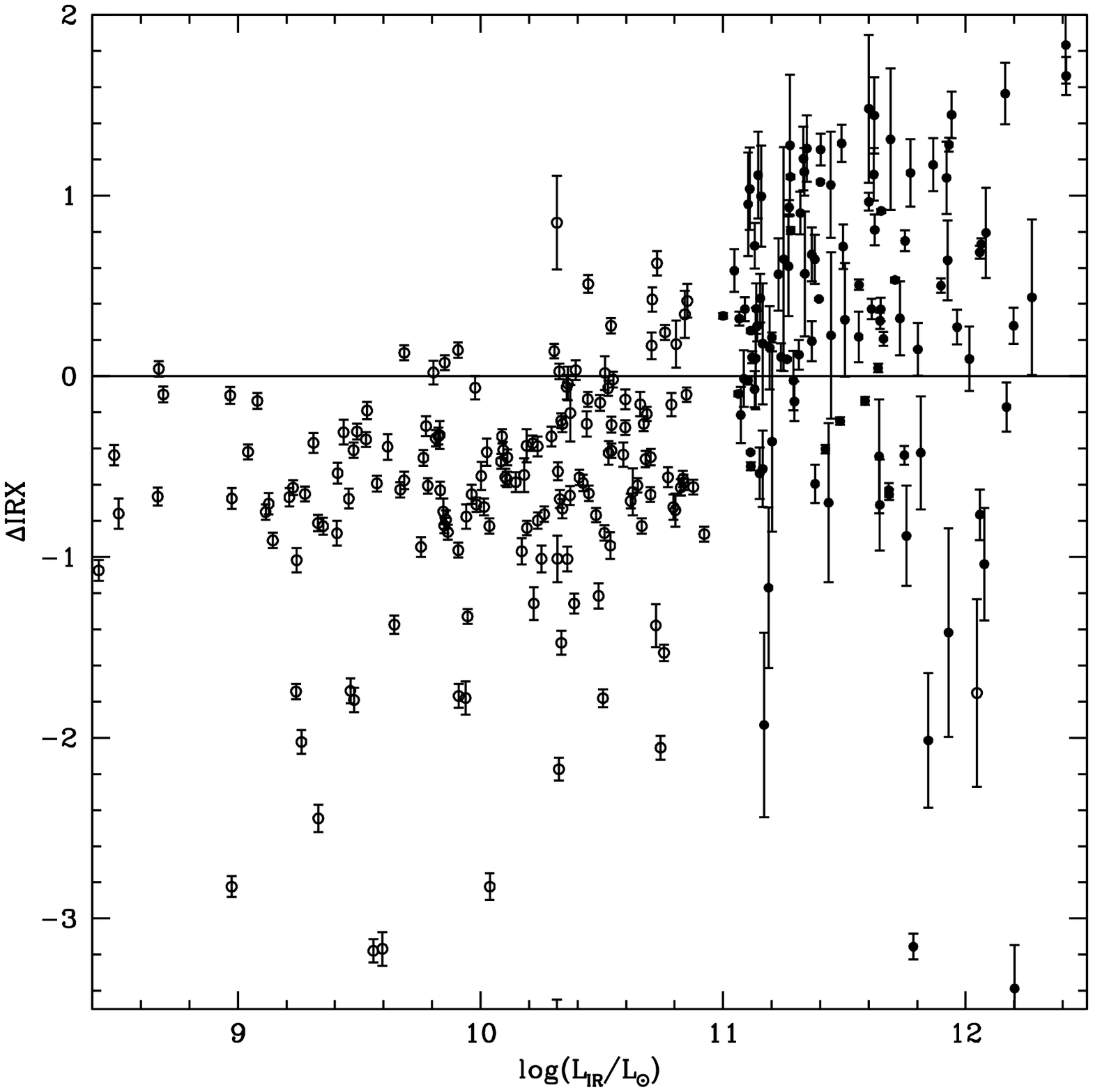}
\figcaption{\small
$\Delta$IRX vs. $L_{IR}$.  GOALS systems are shown as solid points, while 
galaxies from GDP are shown as open points.  $\Delta$IRX increases with IR luminosity 
for $L_{IR}\gtrsim10^{10}~L_{\odot}$.
\label{corr}
}
\end{center}}

\vbox{
\begin{center}
\includegraphics[width=\textwidth]{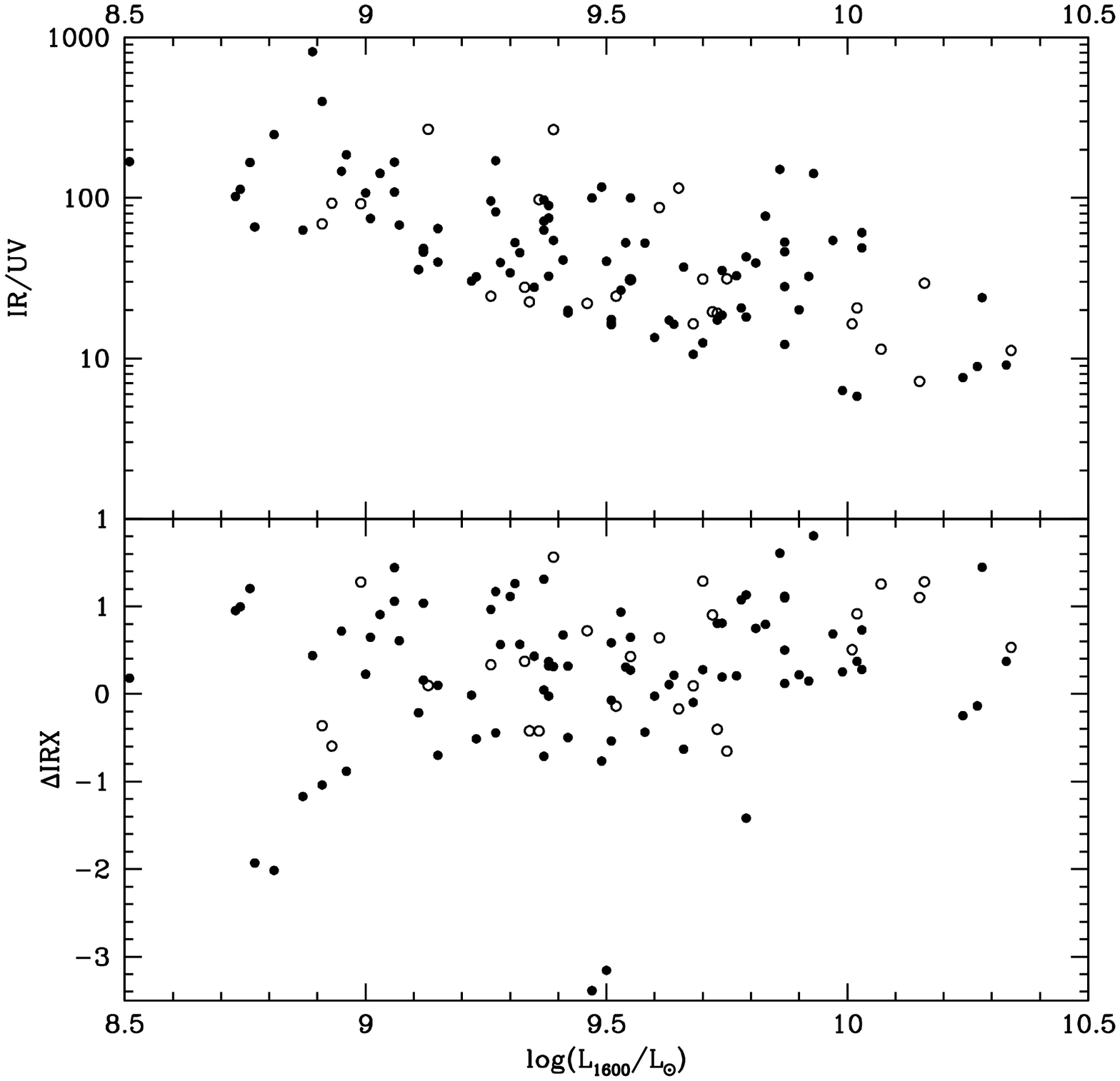}
\figcaption{\small
Top: The IR/UV ratio plotted against $L_{1600}$, the luminosity at $1600{\rm \AA}$ (interpolated 
from FUV and NUV).  The lower envelope shows the sample selection cutoff of 
$L_{IR}>10^{11}~{\rm L}_{\odot}$.
Bottom: $\Delta$IRX vs. $L_{1600}$.  No trend is seen; galaxies of high $\Delta$IRX span 
the full range of UV luminosity.
In both panels, galaxies with IRAC colors suggesting a significant AGN contribution are 
shown as open circles.
\label{l1600}
}
\end{center}}

\vbox{
\begin{center}
\includegraphics[width=\textwidth]{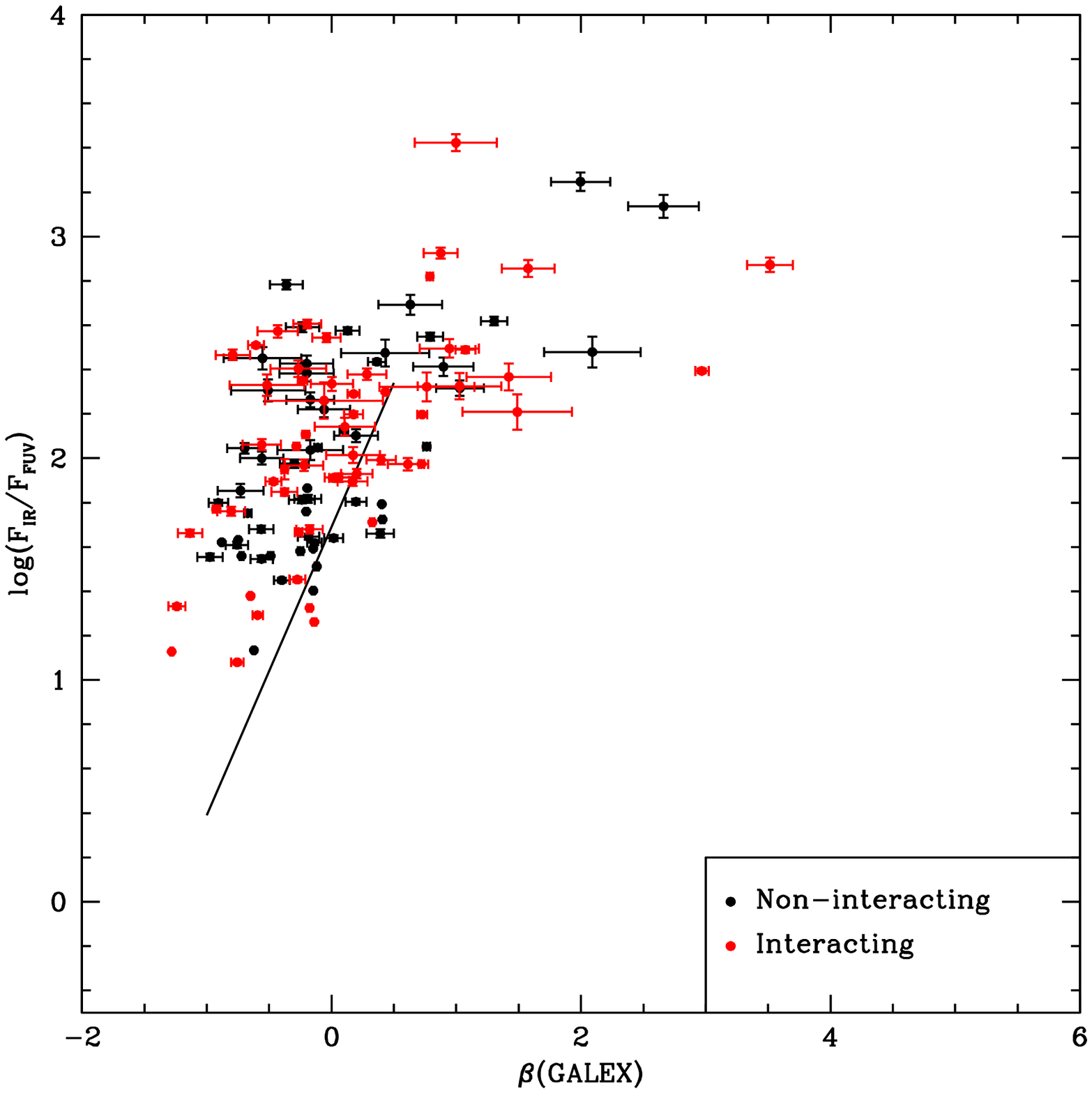}
\figcaption{\small
IRX-$\beta$(GALEX) plot comparing interacting (red) and non-interacting 
(black) LIRG systems.  The solid line is the same as in Fig.~\ref{irxbeta}.  
The interacting and non-interacting populations are consistent with being drawn from 
the same distribution.
\label{merger}
}
\end{center}}

\end{document}